\documentclass[12pt,twocolumn]{article}
\usepackage{fixltx2e}
\usepackage{pictex}
\usepackage{amssymb}
\usepackage{color}
\usepackage{url}
\usepackage[round,sort]{natbib}
\usepackage{fullpage}
\usepackage{graphicx}
\usepackage{float}
\usepackage{abstract}
\usepackage[bf]{caption}
\usepackage{jeep}
\mysection{section}{\large\bf\centering\sc}{\thesection.~}
\newcommand{\gt}[4]{{#1#2 \above 0pt #3#4}}
\newcommand{\gam}[2]{{#1 \above 0pt #2}}
\begin{document}
\title{How Population Growth Affects Linkage Disequilibrium}
\author{Alan R. Rogers\thanks{Dept.{} of Anthropology, 270~S 1400~E,
University of Utah, Salt Lake City, Utah 84112}} 
\twocolumn[
\maketitle
\begin{onecolabstract}
The ``LD curve'' relates the linkage disequilibrium (LD) between pairs
of nucleotide sites to the distance that separates them along the
chromosome. The shape of this curve reflects natural selection,
admixture between populations, and the history of population size. The
present article derives new results about the last of these effects.
When a population expands in size, the LD curve grows steeper, and
this effect is especially pronounced following a bottleneck in
population size. When a population shrinks, the LD curve rises but
remains relatively flat. As LD converges toward a new equilibrium, its
time path may not be monotonic. Following an episode of growth, for
example, it declines to a low value before rising toward the new
equilibrium. These changes happen at different rates for different LD
statistics. They are especially slow for estimates of $\sigma_d^2$,
which therefore allow inferences about ancient population history. For
the European human population, these results suggest a history of
population growth. For several populations of human foragers, they
suggest a history of population decline.

\end{onecolabstract}
]
\saythanks

\section{Introduction}

Linkage disequilibrium (LD) refers to the statistical association
between pairs genetic loci. It is used routinely in localizing disease
genes, in detecting natural selection, and in studying population history.
In all of these contexts, it is necessary to account for the changes
in LD caused by changes in population size.

These changes occur because inhabitants of small populations tend to
be close relatives. The genealogical paths that separate them are
short, and this reduces the opportunity for recombination. For this
reason, LD rises after a fall in population size and falls after a
rise.

These effects are understood in a general way and are often studied by
computer simulation \citep{Kruglyak:NG-22-139, Pritchard:AJH-69-1}.
Although this approach has led to important insights, our
understanding is still rudimentary.  The present paper will use a
deterministic algorithm to explore the effects of growth, of decline,
and of temporary reductions (bottlenecks) in population size. It will
show that each type of history leaves a distinctive signature in the
``LD curve,'' which relates the LD between pairs of sites to the
distance that separates them along the chromosome.

The paper uses $\sigma_d^2$ (defined below) as a measure of LD. This
choice is unusual, because $\sigma_d^2$ has always been of secondary
importance. As we shall see, however, $\sigma_d^2$ has dynamical
properties that give it deeper time depth than alternative measures of
LD. It is readily estimated from data and can be predicted by a
deterministic theory, which makes it easy to study the response to
changes in population size.  This paper will show that $\sigma_d^2$ is
of more than secondary importance. It is useful in its own right as a
measure of LD.

\section{Material and methods}

\subsection{Measuring LD}

Consider a pair of loci (nucleotide sites), $A$ and $B$. At locus $A$,
alleles $A_1$ and $A_0$ have frequencies $a$ and $1-a$. At locus $B$,
alleles $B_1$ and $B_0$ have frequencies $b$ and $1-b$. The
disequilibrium coefficient, $D$, is defined such that $ab + D$ is the
frequency of gamete type $A_1B_1$. The sign of $D$ is arbitrary,
depending on how one labels the alleles, so the magnitude of LD is
often measured by $D^2$.

These measures are strongly affected by heterozygosity at the two
loci, so many authors prefer the squared correlation coefficient
\citep{Hill:TAG-38-226},
\begin{equation}
r^2 = \frac{D^2}{a(1-a)b(1-b)}.
\label{eq.rsq}
\end{equation}
Unfortunately, there is no consensus regarding the expected value of
this statistic, even in the simplest case of neutral loci in a
randomly-mating population of constant size \citetext{compare
  \citealp{Sved:GR-91-183} with \citealp{Song:TPB-71-49} and
  \citealp[p.~98]{Durrett:PMD-08}}.

\citet[p.~233]{Ohta:G-63-229} proposed a related measure of LD, the
``squared standard linkage deviation,'' which was also motivated by a
desire to minimize the effect of heterozygosity.
\begin{equation}
\sigma_d^2 = \frac{E[D^2]}{E[a(1-a)b(1-b)]}
\label{eq.sigdsq}
\end{equation}
This measure is usually viewed as an approximation to $E[r^2]$.  It is
most useful in this role when the population size is large and
constant, and both loci have appreciable heterozygosity
\citep{Hudson:G-109-611}. In other situations, the two parameters can
differ greatly. But even when $\sigma_d^2$ fails as an approximation,
it is still useful as a measure of LD. 

This is not to say that $\sigma_d^2$ provides a complete description
of variation at a pair of loci. That is a problem with multiple
dimensions, which cannot be solved by any scalar-valued measure of LD
\citep[pp.~125--127]{Weir:GDA-96}.  Such measures are simplifications
and are necessarily incomplete \citep{Schaper:G-190-217}. Yet as we
shall see, $\sigma_d^2$ captures enough to provide an interesting
window into the history of population size.

\subsection{Estimation of $\sigma_d^2$}
\label{sec.estimation}

One can estimate $\sigma_d^2$ from data by replacing the expected
values in Eqn.~\ref{eq.sigdsq} with averages across the genome. For
this purpose, I treat each polymorphic site (SNP) as ``focal,'' and
compare the focal SNP with each other SNP within some given
range---often 0.3~cM. From each comparison, I calculate $D^2$ and
$a(1-a)b(1-b)$.  These values are tabulated separately within bins
based on the distance in cM between the SNPs. After all comparisons
have been tabulated, $\hat\sigma_d^2$ is calculated as the sum of
$D^2$ within a bin divided by the corresponding sum of $a(1-a)b(1-b)$.

To estimate uncertainties, I use a moving blocks bootstrap
\citep{Lieu:ELB-92-225}, with 300 SNPs per block.  With diploid data,
some genotypes may be unphased. In such cases, I use the EM algorithm
to estimate $D^2$, as explained in the appendix.

\subsection{Deterministic evolution of $\sigma_d^2$}
\label{sec.hill}

Under neutral evolution with constant population size, $\sigma_d^2$
converges to an equilibrium value \citep{Ohta:G-68-571,
  McVean:G-162-987}. In addition, several authors have introduced
recurrence equations, which make it possible to study the transient
behavior of $\sigma_d^2$ after changes in population size
\citep{Weir:TPB-6-323, Hill:TPB-8-117, Strobeck:G-88-829}. The model
of \citet{Strobeck:G-88-829} allows faster calculations but is less
numerically stable than that of \citet{Hill:TPB-8-117}. I present some
results using the former model of but focus primarily on the latter.
In the appendix, I summarize Hill's model and show that it holds not
only under the mutational model that he studied, but also under the
model of infinite sites \citep{Kimura:InfiniteSites}.  It is thus
appropriate for use with DNA sequence data.

For bootstrap confidence intervals, I accelerate these calculations by
using Eqn.~\ref{eq.DRMx} of the appendix to approximate a system of
ordinary differential equations, which is then solved using standard
software.

\subsection{Simulations}

The various methods were evaluated against simulated data generated by
MACS \citep{Chen:GR-19-136}. Simulation results in Fig.~\ref{fig.TH}
are based on my own coalescent program, written in C.

\subsection{Sampling bias}

Drawing a sample is equivalent to one generation of evolution with a
very small population---one whose size equals the sample size. Because
drift introduces LD, there is much more LD in a sample than in the
population from which it was drawn.

\citet[Eqn.~6]{Hudson:G-109-611} showed that sampling bias in $r^2$
equals $1/n$, when the sample consists of $n$ gametes. To correct
sampling bias in $\hat\sigma_d^2$, I use the model of
\citet{Hill:TPB-8-117} or \citet{Strobeck:G-88-829} to project the
state vector forward one generation, with the population size set
equal to the sample size.

\section{Results}

\subsection{$\hat\sigma_d^2$ is average $r^2$, weighted by  heterozygosities}
\label{sec.Hweighting}

Eqn.~\ref{eq.sigdsq} is equivalent to
\begin{equation}
\sigma_d^2 = \frac{E[H_A H_B r^2]}{E[H_A H_B]},
\label{eq.sigdsq.weighted}
\end{equation}
where $H_A = 2a(1-a)$ is the heterozygosity at locus $A$ and $H_B$ is
that at locus $B$. This implies that $\sigma_d^2$ is the expectation
of $r^2$ weighted by the product of the two heterozygosities.

This weighting also carries over to the estimate, $\hat\sigma_d^2$,
which is obtained by replacing expectations with averages. Such
estimates will be insensitive to loci with low heterozygosity and, for
this reason, also insensitive to sequencing error. They should be
useful with low-coverage sequence data.

\begin{figure*}
{\centering\input{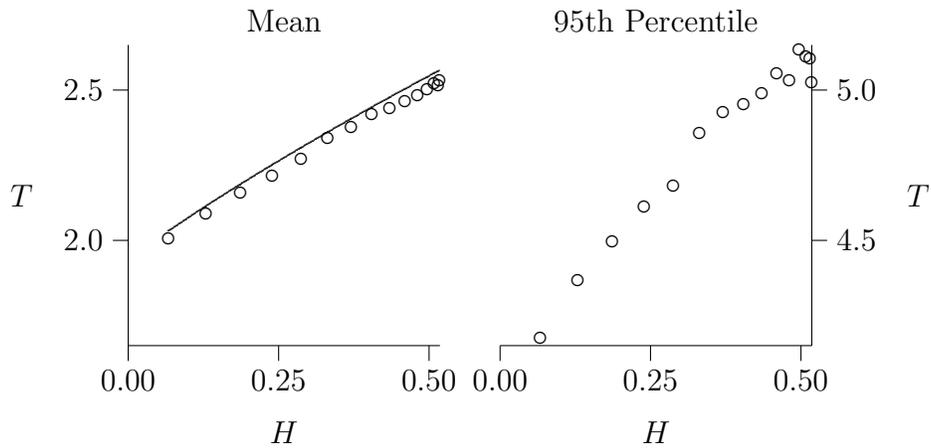}\\}
\caption{\textbf{The depth, $T$, of gene genealogy given heterozygosity,
  $H$.} The left and right panels show the mean and the 95th
  percentile. Solid line shows expected values,
  based on the model of \citet[Eqn.~1.5,
  p.~276]{Griffiths:SM-14-273}. Open circles show results from
  coalescent simulations. These results assume a sample of 30
  individuals and a mutation rate of $0.02$ per $2N$ generations. The
  slope is greater in the right panel than the left, indicating that
  heterozygosity has a stronger effect on the upper tail of the
  distribution than on the mean.}
\label{fig.TH}
\end{figure*}

\subsection{Loci with high heterozygosity have deep gene trees.}
\label{sec.deeptree}
Weighting by heterozygosity is also of interest because it exaggerates
the influence of unusually deep gene trees. At some given nucleotide
position, let $T$ represent the age of the last common ancestor of the
sample. I will call this the ``depth'' of the gene tree for that
nucleotide position. \citet[Eqn.~1.5, p.~276]{Griffiths:SM-14-273}
derive the conditionally expected depth, $E[T|x,n]$, given that $x$
copies of the derived allele were observed in a sample of haploid size
$n$. Given the heterozygosity, we can solve for $x$, but we cannot
tell the derived from the ancestral allele. It may be present in $x$
copies or in $n-x$. At mutation-drift equilibrium, however, these
alternatives have probabilities proportional to $1/x$ and $1/(n-x)$
\citep[Eqn.~1]{Fu:TPB-48-172}. Using these values as weights, I
average $E[T|x,n]$ and $E[T|n-x,n]$ in order to calculate expected
tree depth given heterozygosity. The results are shown as a solid line
in Fig.~\ref{fig.TH}. As the figure shows, tree depth increases with
heterozygosity. Simulated values---shown as open circles---agree
closely with the theory.

Simulated values also allow us to examine the upper tail of the
distribution, as seen in the right panel of Fig.~\ref{fig.TH}.  The
95th percentile of tree depth increases with heterozygosity even more
steeply than does the mean. Thus, heterozygosity has an exaggerated
effect on the upper tail of the distribution.

Geneticists often study loci selected for their high heterozygosity
\citep{Lewontin:AJH-19-681, Rogers:AJH-58:1033, Clark:GR-15-1496}. A
sample of loci selected in this fashion will have gene trees that are
unusually deep, especially in the upper tail of the distribution.

Because $\sigma_d^2$ is weighted by heterozygosity, it exaggerates the
influence of these deep gene trees. For this reason, it should be
sensitive to earlier portions of a population's history. Following a
change in population size, it should converge more slowly to the new
equilibrium.

\begin{figure*}
{\centering\input{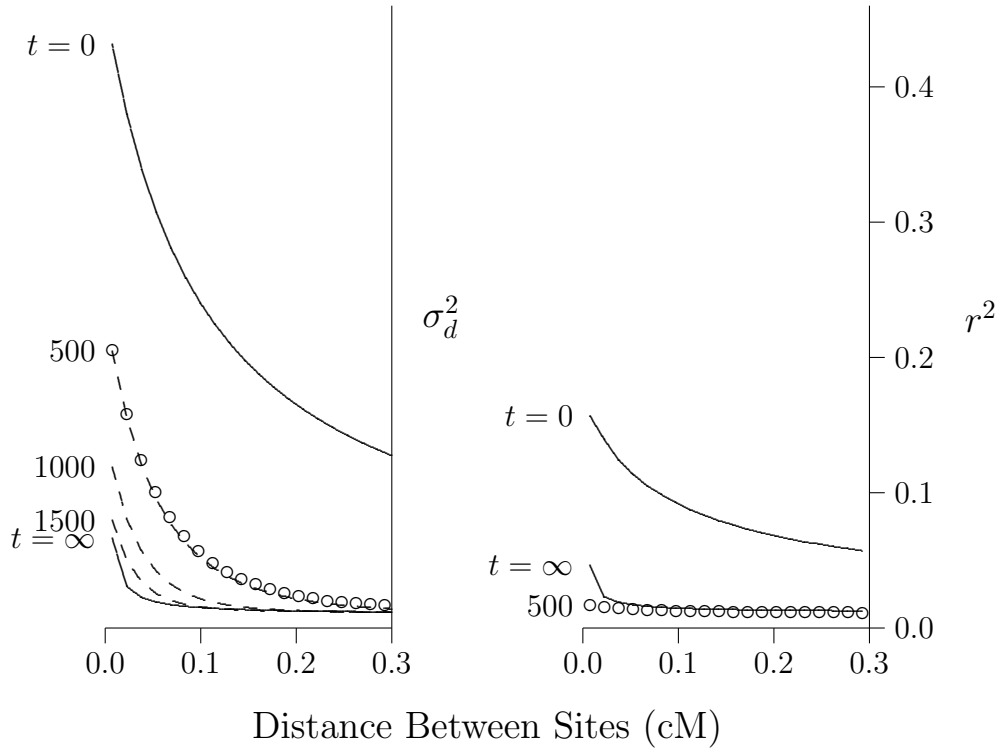}\\}
\caption{\textbf{Effect of population expansion on the LD curve.}
  The population grew suddenly at time $t=0$ from $2N=10^3$ to $10^5$.
  \emph{Left panel:} LD is measured by $\sigma_d^2$. Solid lines show the
  predicted values at the initial equilibrium ($t=0$) and at the
  eventual equilibrium $(t=\infty)$. Dashed lines show a series of transient
  states that occur at various values of $t$, the number of
  generations since the expansion. These lines are all calculated
  using the method of \citet{Hill:TPB-8-117}. Open circles show
  values simulated 
  using MACS \citep{Chen:GR-19-136} for $t=500$.
  \emph{Right panel:} LD is measured by $r^2$, and points and lines are
  based on computer simulation. The mutation rate is $u=10^{-8}$ per
  site per generation, and the haploid sample size is 100.}
\label{fig.gro}
\end{figure*}

\subsection{Effect of population growth}
\label{sec.growth}

LD will decline following an expansion of population size, because
genetic drift is weaker in large populations and produces less
LD. This process is illustrated in Fig.~\ref{fig.gro}, which shows the
effect of an expansion from size $2N=10^3$ to $2N=10^5$. The LD curve
of the initial population (labeled $t=0$), represents an equilibrium
between mutation, drift, and recombination. After the expansion, the
LD curve will eventually converge to a new equilibrium, which is
labeled $t=\infty$. This new equilibrium, however, is reached only
gradually.

LD is measured by $\sigma_d^2$ in the left panel and by $r^2$ in the
right. The dashed lines were calculated using the method of
\citet{Hill:TPB-8-117}, and the open circles were estimated from data
simulated using MACS \citep{Chen:GR-19-136}. Note that $\sigma_d^2$ is
still far from equilibrium at generation~500 and does not approach
equilibrium until about generation 1500.

The situation is very different in the right panel, where LD is
measured using $r^2$. By generation~500, $r^2$ is close to
equilibrium. At the left edge of the graph, it is slightly
\emph{below} the equilibrium. We'll return to this point later, but
for the moment, note simply that $r^2$ converges much faster than
$\sigma_d^2$. For this reason, the two measures are useful for
studying different time scales. We learn from $r^2$ about the recent
past and from $\sigma_d^2$ about more ancient events. Presumably, this
difference arises because $\sigma_d^2$ is weighted toward loci with high
heterozygosity. As shown in Fig.~\ref{fig.TH}, this weighting makes it
more sensitive to the distant past.

Returning to the left panel of Fig.~\ref{fig.gro}, note that the right
portion of the curve converges faster than the left. This is because
the post-expansion population is large, and genetic drift is weak. The
dynamics are therefore dominated by recombination, which is stronger
on the right side of the graph. The result is that midway through the
process---say at generation~500---the LD curve declines very
steeply. Thus, a steeply declining LD curve is the signature of recent
population growth. Steeply declining LD curves can also be produced by
gene conversion \citep{Frisse:AJH-69-831}.  The two effects should be
separable, however, because population growth affects LD over a much
wider range of recombination rates. The effect of population growth on
the LD curve was first described by \citet{Kruglyak:NG-22-139}.

\subsection{The method of \citeauthor{Hayes:GR-13-635}}
\label{sec.hayes}

Figure~\ref{fig.gro} suggests that $\sigma_d^2$ is likely to be of
greater use than $r^2$ in inferring the ancient history of population
size. Yet the latter statistic is often used instead
\citep{Hayes:GR-13-635, Tenesa:GR-17-520, McEvoy:GR-21-821}, using a
method that is based on the formula $E[r^2] \approx \rho$, where
\begin{equation}
\rho = 1/(1 + 4Nc)
\label{eq.sved}
\end{equation}
Here, $N$ is population size, and $c$ is the recombination rate
\citep{Sved:TPB-2-125, Sved:TPB-4-129}.  Although this formula has
been criticized \citetext{\citealp[p.~272]{Littler:TPB-4-259};
  \citealp[pp.~987--988]{McVean:G-162-987};
  \citealp[p.~98]{Durrett:PMD-08}}, it is still used as a basis for
inference.

\citet{Hayes:GR-13-635} study a generalization of $r^2$ and argue from
simulations that the expectation of this statistic is approximately
equal to $\rho$.  Furthermore, this approximation even works for
populations that have increased in size at a constant linear rate,
provided that one interprets $N$ as the population size that obtained
$1/2c$ generations in the past.  By inverting Eqn.~\ref{eq.sved}, they
are able to estimate $N$ over a wide range of values in $c$, which
correspond under their model to different points in the past.
\citet{Tenesa:GR-17-520} work directly with $r^2$, but estimate the
history of population size in the same way. This method is also used
by \citet{McEvoy:GR-21-821}.

Of course, no population can increase linearly forever, so the period
of linear increase must end at some point in the past. Furthermore,
the method has also been used to infer more complex patterns of
growth, including bottlenecks \citep{Hayes:GR-13-635,
  Tenesa:GR-17-520, McEvoy:GR-21-821}.  Thus, it is worth evaluating
the method in a broader context.

\begin{figure*}
{\centering\input{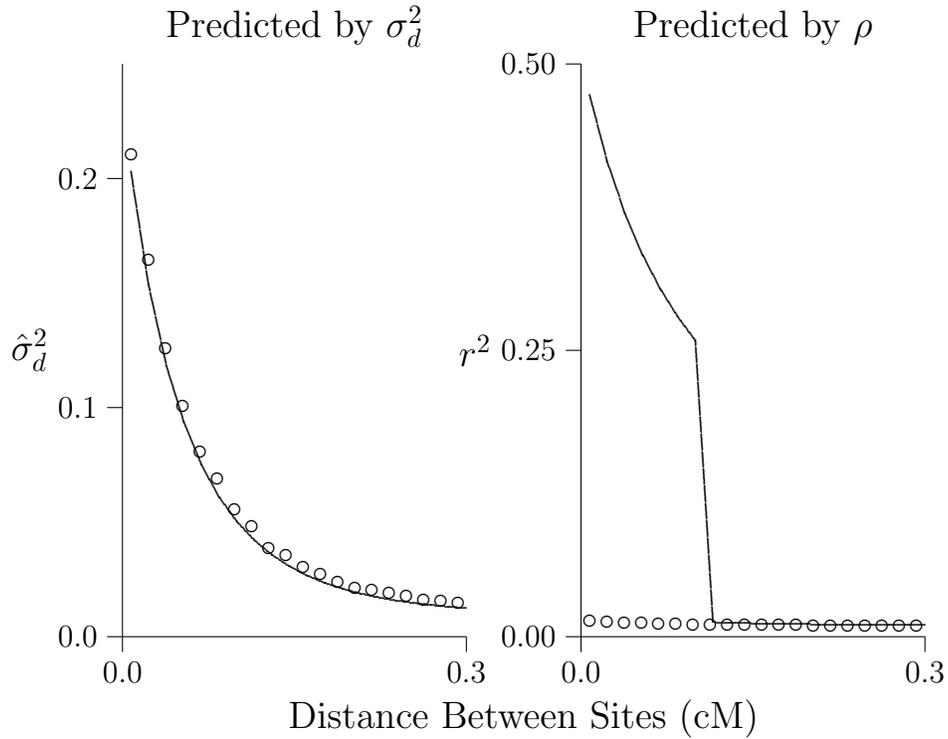}\\}
\caption{\textbf{Predicted and simulated LD.} Each panel shows the
  same population 500 generations after expansion from $2N=10^3$ to
  $10^5$. Lines show values predicted by $\sigma_d^2$ (left panel,
  \citep{Hill:TPB-8-117}) and by $\rho$ (right panel,
  \citep{Hayes:GR-13-635}). Open circles show results estimated from
  simulations.}
\label{fig.hillhayes}
\end{figure*}

\begin{figure*}
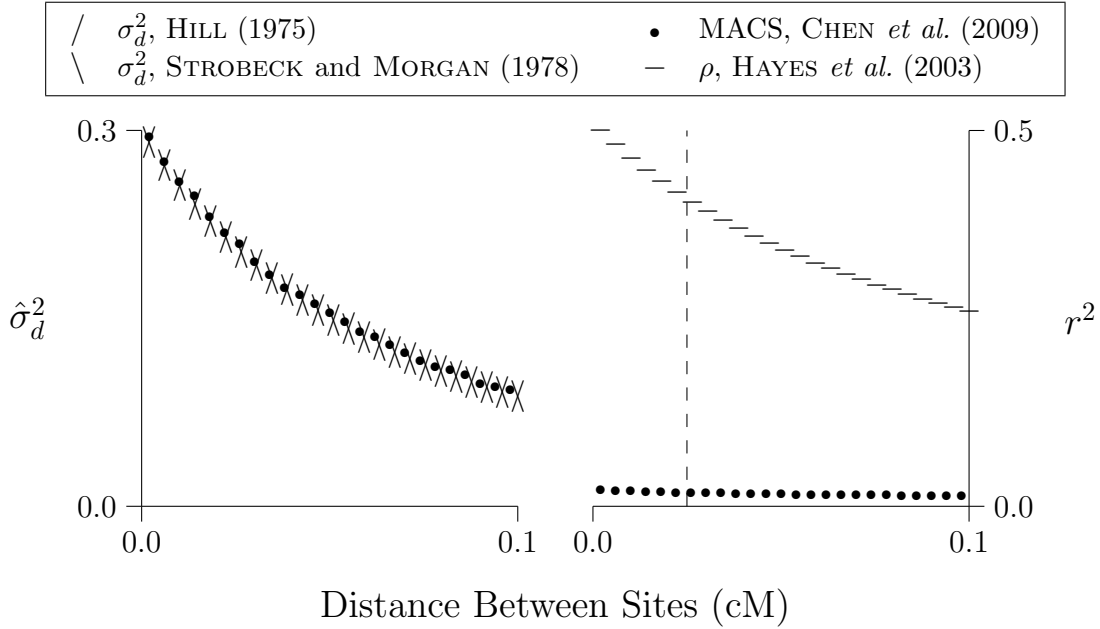

{\centering 
%
\mbox{\beginpicture
  \setcoordinatesystem units <50cm,16.67cm> point at 0.11 0
  \setplotarea x from 0 to 0.1, y from 0 to 0.3
  \axis left label {\large $\hat\sigma_d^2$} 
  ticks numbered from 0.0 to 0.3 by 0.3 /
  \axis bottom ticks numbered from 0.0 to 0.1 by 0.1 /
  \put {\large Distance Between Sites (cM)} [tc] <0pt,-6ex> at 0.11 0
  \put{\fbox{\begin{tabular}{cl@{\hspace{2em}}cl}
      $/$ & \small $\sigma_d^2$, \citet{Hill:TPB-8-117}
      & \scriptsize$\bullet$ & \small MACS, \citet{Chen:GR-19-136}\\
      $\backslash$ & \small $\sigma_d^2$, \citet{Strobeck:G-88-829}
      & $-$ & \small $\rho$, \citet{Hayes:GR-13-635}\\
  \end{tabular}}} [b] <0pt,2.5ex> at 0.11 0.3
\multiput {\scriptsize$\bullet$} at
 0.00200031  0.29435870
 0.00599947  0.27463653
 0.01000024  0.25872866
 0.01399998  0.24760212
 0.01799988  0.23061433
 0.02199960  0.21781900
 0.02600009  0.20887413
 0.02999914  0.19519601
 0.03399954  0.18441224
 0.03800137  0.17430551
 0.04200035  0.16874650
 0.04599895  0.16092769
 0.04999969  0.15401227
 0.05400062  0.14660886
 0.05799927  0.13900402
 0.06199882  0.13464251
 0.06600182  0.12883103
 0.07000062  0.12214023
 0.07399808  0.11608463
 0.07800095  0.11107957
 0.08199865  0.10857199
 0.08600060  0.10430120
 0.09000048  0.09769103
 0.09399846  0.09517354
 0.09800004  0.09273960
/
\multiput {$/$} at
0.002000000 0.290566996
0.006083333 0.272610730
0.010166667 0.256231733
0.014250000 0.241244613
0.018333333 0.227491577
0.022416667 0.214837489
0.026500000 0.203165950
0.030583333 0.192376169
0.034666667 0.182380436
0.038750000 0.173102075
0.042833333 0.164473766
0.046916667 0.156436165
0.051000000 0.148936762
0.055083333 0.141928931
0.059166667 0.135371128
0.063250000 0.129226226
0.067333333 0.123460943
0.071416667 0.118045360
0.075500000 0.112952514
0.079583333 0.108158038
0.083666667 0.103639865
0.087750000 0.099377962
0.091833333 0.095354102
0.095916667 0.091551670
0.100000000 0.087955490
/
\multiput {$\backslash$} at
0.002000000 0.290450785
0.006083333 0.272465623
0.010166667 0.256064225
0.014250000 0.241077608
0.018333333 0.227311562
0.022416667 0.214655803
0.026500000 0.202976486
0.030583333 0.192169538
0.034666667 0.182164177
0.038750000 0.172892421
0.042833333 0.164259195
0.046916667 0.156220950
0.051000000 0.148720199
0.055083333 0.141715226
0.059166667 0.135153007
0.063250000 0.129011347
0.067333333 0.123242510
0.071416667 0.117832749
0.075500000 0.112734387
0.079583333 0.107945190
0.083666667 0.103422230
0.087750000 0.099165000
0.091833333 0.095135898
0.095916667 0.091339925
0.100000000 0.087751897
/
  \setcoordinatesystem units <50cm,10cm> point at -0.01 0
  \setplotarea x from 0 to 0.1, y from 0 to 0.5
  \axis right label {\large $r^2$} 
  ticks numbered from 0.0 to 0.5 by 0.5 /
  \axis bottom ticks numbered from 0.0 to 0.1 by 0.1 /
\setdashes
\putrule from 0.025 0 to 0.025 0.5
\setsolid
\multiput {$-$} at
0.002000000 0.500196078
0.006083333 0.481327573
0.010166667 0.463857791
0.014250000 0.447636761
0.018333333 0.432535211
0.022416667 0.418441116
0.026500000 0.405256917
0.030583333 0.392897256
0.034666667 0.381287129
0.038750000 0.370360360
0.042833333 0.360058343
0.046916667 0.350328985
0.051000000 0.341125828
0.055083333 0.332407308
0.059166667 0.324136126
0.063250000 0.316278714
0.067333333 0.308804781
0.071416667 0.301686923
0.075500000 0.294900285
0.079583333 0.288422274
0.083666667 0.282232305
0.087750000 0.276311585
0.091833333 0.270642919
0.095916667 0.265210549
0.100000000 0.260000000
/
\multiput {\scriptsize$\bullet$} at
 0.00200031  0.02158001
 0.00599947  0.02048985
 0.01000024  0.01999984
 0.01399998  0.01929721
 0.01799988  0.01875431
 0.02199960  0.01808524
 0.02600009  0.01779525
 0.02999914  0.01744106
 0.03399954  0.01690631
 0.03800137  0.01649251
 0.04200035  0.01635699
 0.04599895  0.01605933
 0.04999969  0.01573412
 0.05400062  0.01542379
 0.05799927  0.01526753
 0.06199882  0.01506781
 0.06600182  0.01482748
 0.07000062  0.01456646
 0.07399808  0.01450749
 0.07800095  0.01415418
 0.08199865  0.01412382
 0.08600060  0.01392211
 0.09000048  0.01368248
 0.09399846  0.01351872
 0.09800004  0.01344594
/
\endpicture}
\\}
\caption{\textbf{Linkage disequilibrium ($\hat\sigma_d^2$ or $r^2$)
    after an episode of linear growth.} The population begins at
  mutation-drift equilibrium with $2N=10^3$ and then grows linearly to
  $2N=10^6$ over a period of 300 generations. Linear growth is
  approximated by 20 epochs of 15 generations, within each of which
  $2N$ is constant.  In the right panel, the region to the right of
  the vertical line corresponds to the period of linear growth,
  according to the model \citep{Hayes:GR-13-635}.  All calculations
  assume that $u=1.48\times 10^{-8}$ and $c=10^{-8}$ per
  nucleotide. Simulations involve $10^9$ base pairs of DNA, which are
  sequenced in a sample of 100 homologous chromosomes.}
\label{fig.lingro}
\end{figure*}

Figure~\ref{fig.hillhayes} considers a population that expands
suddenly in size and is observed 500 generations later.  In both
panels, the solid lines show predicted LD. In the left panel, LD is
measured by $\hat\sigma_d^2$ and predicted by $\sigma_d^2$
\citep{Hill:TPB-8-117}. In the right panel, it is measured by $r^2$
and predicted by $\rho$ \citep{Hayes:GR-13-635}. The open circles show
values simulated using MACS \citep{Chen:GR-19-136}.  In the context of
this population history, $\sigma_d^2$ provides excellent predictions
but $\rho$ does not.

Figure~\ref{fig.lingro} evaluates $\rho$ against a history involving
300 generations of linear population growth, approximated as a series
of 20 steps. Linear growth is the context in which the model was
originally validated, so it ought to work well here.  Yet this is not
the case. The period of linear growth corresponds to the region to the
right of the dashed line in the right panel of
Fig.~\ref{fig.lingro}. At least in this region, predicted and
simulated values should match. Yet as before, $\sigma_d^2$ predicts
well, $\rho$ poorly.

\begin{figure}
{\centering\input{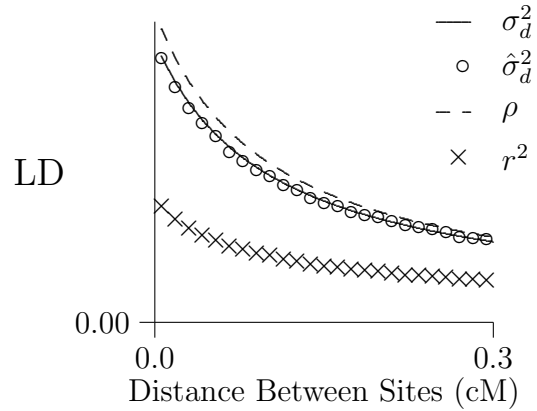}\\}
\caption{\textbf{LD curve under constant population size.} Simulated
  values of $r^2$ and $\hat\sigma_d^2$ were generated using MACS
  \citep{Chen:GR-19-136}.  $\sigma_d^2$ was calculated using the
  method of \citet{Hill:TPB-8-117}, and $\rho$ was calculated from
  Eqn.~\ref{eq.sved}. All calculations assume a population size of
  $2N=10^3$ and a sample of 60 chromosomes.}
\label{fig.const} 
\end{figure}

Finally, Fig.~\ref{fig.const} evaluates $\rho$ and $\sigma_d^2$
against a history of constant population size. In this case, $\rho$ is
a fair approximation to $\hat\sigma_d^2$ but does not provide a useful
approximation of $r^2$. On the other hand, $\sigma_d^2$ does provide
an accurate prediction of $\hat\sigma_d^2$.

There may be cases in which $\rho$ predicts $r^2$ with accuracy, but
this is not so in any of the cases studied here. It seems unlikely
that $\rho$ will be useful as a basis for inference in real
populations. 

\begin{figure}
{\centering\input{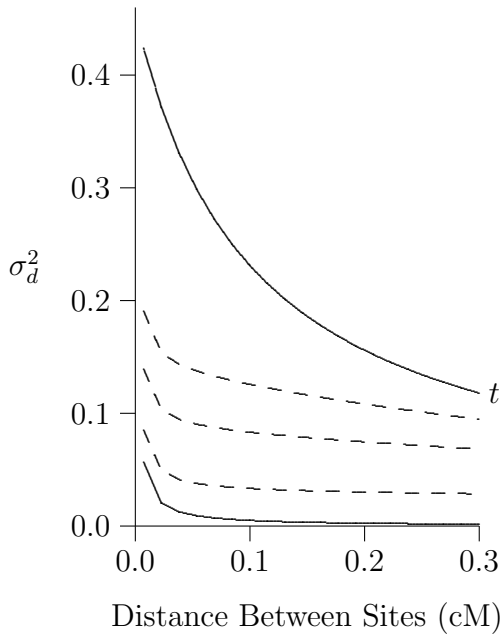}\\}
\caption{\textbf{Effect of population collapse on the LD curve.} At
  time $t=0$, this population collapsed in size from $2N=10^5$ to
  $10^3$. Other details are as in Fig.~\ref{fig.gro}.}
\label{fig.shrink}
\end{figure}

\subsection{Effect of population collapse}
\label{sec.collapse}

A collapse in population size produces an effect on the LD curve very
different from that of population growth. This is shown in
Fig.~\ref{fig.shrink}, which illustrates two important points.  First,
the whole process is over very quickly. Even with $\sigma_d^2$, we
cannot look very far into the past.  Second, the transient curves (the
dashed lines) are quite flat. As time passes, the initial curve (at
$t=0$) is elevated without much change in shape. Presumably, this
is because the effect of drift is dominant in a small population, so
that differences in recombination rate do not matter much. The result
is that the LD curve becomes high but flat after a collapse in
population size.

\begin{figure}
{\centering\input{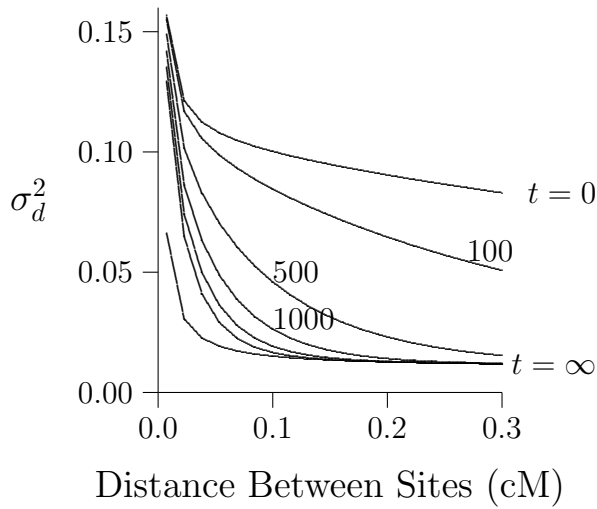}\\}
\caption{\textbf{Effect of a 100-generation bottleneck in population
    size.} The curves show $\sigma_d^2$ at various points after
  recovery from a 100-generation bottleneck, during which the
  population had size $2N=10^3$. Before and after the bottleneck, its
  size was $10^5$. Other details as in Fig.~\ref{fig.gro}.}
\label{fig.bottle}
\end{figure}

\subsection{Effect of a bottleneck}
\label{sec.bottle}

Figure~\ref{fig.bottle} shows the effect of a 100-generation
bottleneck in population size. Before and after the bottleneck, this
population had size $2N=10^5$. During the bottleneck, its size was
$10^3$. The bottleneck ended at time~0, and we observe it in various
subsequent generations. In the graph, the curve labeled $t=0$ is at
the end of the bottleneck and exhibits an LD curve that is high but
flat, for reasons discussed in the preceding paragraph. As time
passes, the right portion of the curve (the portion with high rates of
recombination) falls much faster than the left, so that by
generation~1000, the curve is almost L-shaped. This is the signature
of a bottleneck in population size.

Why should there be such a difference between expansion from an
initial equilibrium and from a bottleneck?---between
Figs.~\ref{fig.gro} and~\ref{fig.bottle}? Both curves are declining
toward the same equilibrium, but the left portion of the curve
declines much more slowly after a bottleneck than after expansion from
equilibrium. This can only reflect the state of the initial
population, just before the increase in size. The initial population
of Fig.~\ref{fig.gro} had been small much longer than that of
Fig.~\ref{fig.bottle}. Consequently, it had less heterozygosity. As we
shall see in the next section, this accelerates the rate of decline.

\begin{figure}
{\centering\input{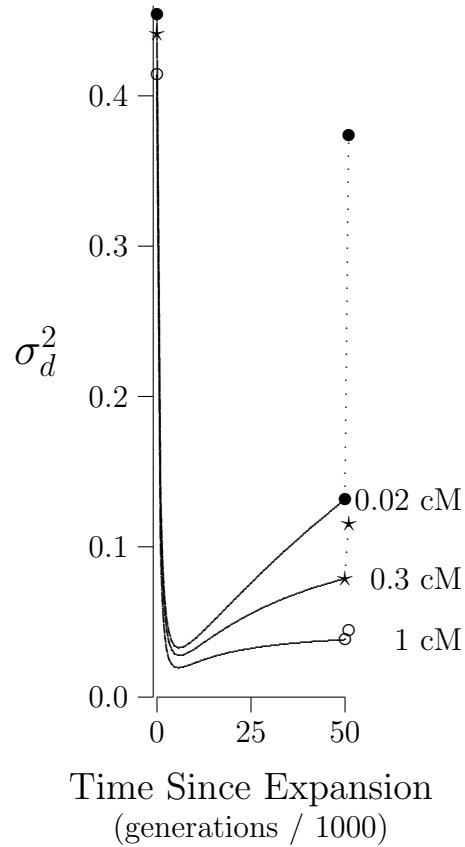}\\}
\caption{\textbf{The time path of $\sigma_d^2$ after a population
    expansion.} The initial population was small ($2N=10^3$) and at
  mutation-drift equilibrium. At time~0 it expands suddenly to a much
  larger size ($2N = 10^5$). Each curve shows the time path of
  $\sigma_d^2$ for a different rate of recombination, measured in
  centimorgans (cM). There is a curve for tight linkage ($\bullet$,
  0.02~cM), one for somewhat weaker linkage ($\star$, 0.3~cM), and one
  for linkage that it weaker still ($\circ$, 1~cM). Dotted lines
  connect the final value of each curve to its ultimate equilibrium,
  which would be reached if the population stayed large forever.}
\label{fig.tpath}
\end{figure}

\subsection{The time paths of individual points on the LD curve}
\label{sec.tpath}

Figure~\ref{fig.tpath} shows the time path of $\sigma_d^2$ after an
expansion in population size. Because the new population is larger,
the new equilibrium will have less LD. But $\sigma_d^2$ does not
decline monotonically toward this new equilibrium. Instead, it falls
rapidly to a smaller value before climbing back slowly towards to the
new equilibrium.

\begin{figure}
{\centering\input{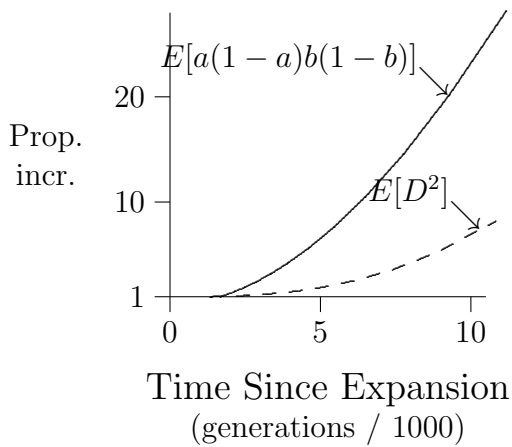}\\}
\caption{\textbf{The time path of numerator and denominator of $\sigma_d^2$
  following a population expansion.}  Dashed line shows numerator of
  $\sigma_d^2$, relative to its value in generation~1000. Solid line
  shows the denominator in the same way. Loci are separated by
  0.02~cM. Population history is as in Fig.~\ref{fig.tpath}.}
\label{fig.numden}
\end{figure}

The initial decline in $\sigma_d^2$ does not result from any decline
in its numerator, $E[D^2]$.  Indeed, Fig.~\ref{fig.numden} shows that
the numerator actually grows. The decline in $\sigma_d^2$ happens
because growth in its numerator is outstripped by that in its
denominator, $E[a(1-a)b(1-b)]$, which increases under the influence of
mutation.  The proportional increase is large, because our initial
population that was small and therefore had little heterozygosity. For
this reason, the denominator was initially so small that increments
caused by mutation had a large proportional effect.

Two factors account for the post-expansion growth in $E[D^2]$, the
numerator of $\sigma_d^2$.  First, $D$ is proportional to
heterozygosity \citep[p.~334]{Kaplan:AJH-51-333}, which increases in
response to mutation. Ordinarily, the positive effect on $D^2$ would
be offset by the negative effect of recombination. But in the early
generations following our expansion, this negative effect is very
weak. This is because the initial population was small and thus had
low heterozygosity. Few recombinant gametes are produced in such a
population, because such gametes are produced only by double
heterozygotes. Consequently, the effect of recombination is weaker
than that of mutation.

The non-monotone time path in Fig.~\ref{fig.tpath} refers to
$\sigma_d^2$, but it seems plausible that $r^2$ might obey similar
dynamics. This may explain why, in the right panel of
Fig.~\ref{fig.gro}, the left end of the curve for $t=500$ is below the
equilibrium.

This also explains why the left portion of the LD curve declines
faster after expansion from equilibrium than after a bottleneck.  The
left portion of the curve refers to tightly-linked sites, with weak
recombination. In the post-expansion population, genetic drift is also
weak because the population is large. Because recombination and drift
are both weak, mutation dominates the dynamics. The proportional
effect of mutation is large when heterozygosity is low. In the case of
expansion from equilibrium, the population has been small a long time,
so heterozygosity is low, the proportional effect of mutation is
large, and $\sigma_d^2$ declines rapidly. Heterozygosity is not so low
at the end of a bottleneck, because the population has been small only
briefly. Consequently, the proportional effect of mutation is smaller,
and $\sigma_d^2$ declines more slowly.

\begin{figure}
{\centering\includegraphics[width=\columnwidth]{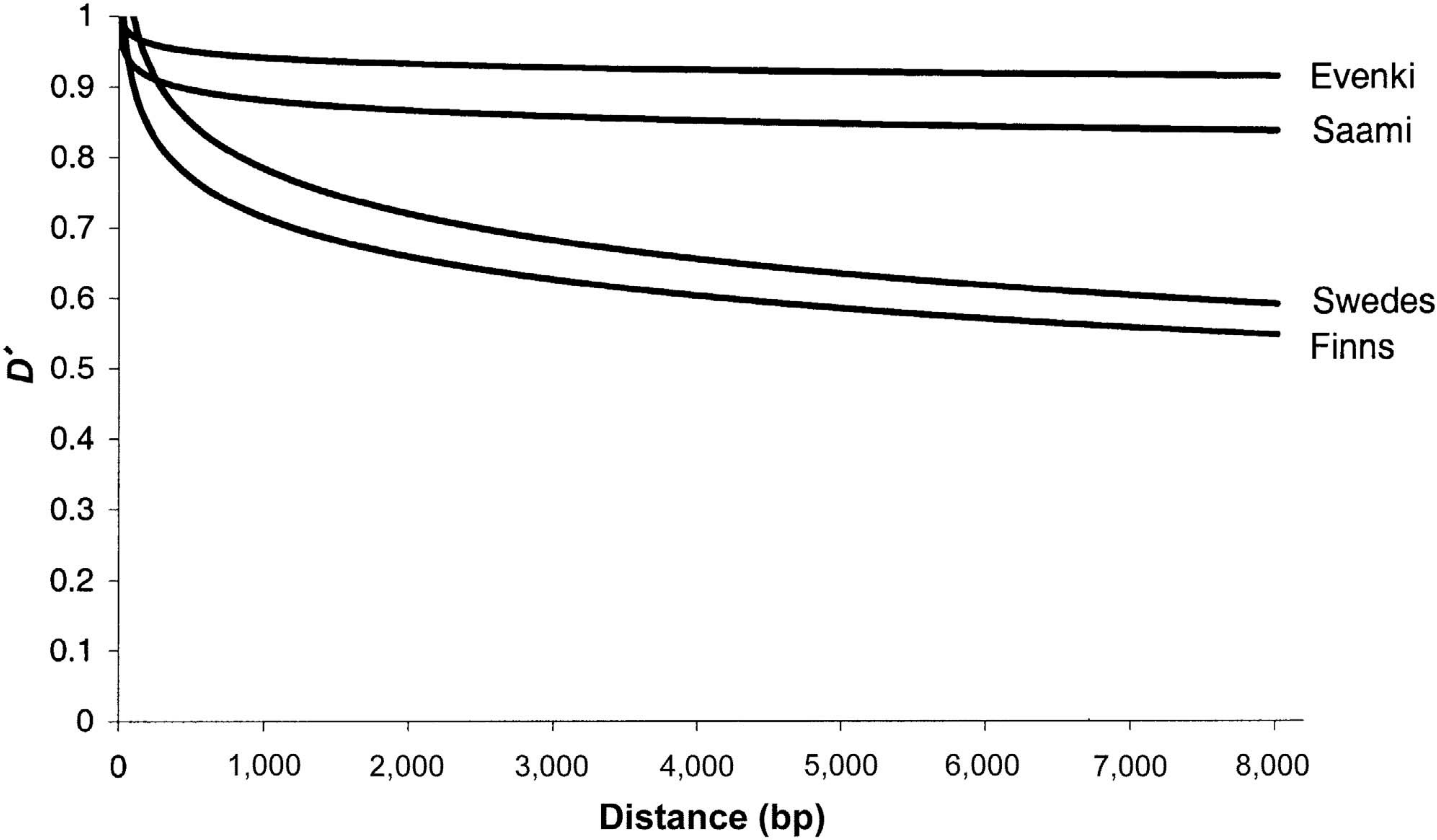}\\}
\caption{\textbf{LD curves for four human populations.} Source:
  \citet[Fig.~3]{Kaessmann:AJH-70-673}.} 
\label{fig.kaessmann}
\end{figure}

\subsection{LD in real populations}
\label{sec.realpop}

When applied to data from real populations, the present methods may
suggest one history or another. Such suggestions can only be
tentative, because they take no account of statistical
uncertainty. Nonetheless, they may be useful in exploratory data
analysis. In this spirit, let us examine the LD curves of several
human populations.

Figure~\ref{fig.kaessmann} shows LD curves published by
\citet{Kaessmann:AJH-70-673} for four populations. They measure LD
using the statistic $D'$, which is a normalized variant of $D$
\citep[p.~55]{Lewontin:G-49-49}. Two of the curves---those for the
Evenki and the Saami---refer to small high-latitude foraging
populations. The other two refer to the large populations of nation
states. The curves of the two foraging populations are not only high,
but also much flatter than those of the nation states. They resemble
the dashed lines in
Fig.~\ref{fig.shrink}. \citet[p.~681]{Kaessmann:AJH-70-673} attribute
elevated LD in the foraging populations to a long history of constant
size. The present analysis, however, suggests a different
interpretation.  If $D'$ behaves like $\hat\sigma_d^2$, the flat LD
curves suggest that both foraging populations have recently declined
in size. On the basis of similar data,
\citet[p.~7]{Pritchard:AJH-69-1} suggest a third hypothesis---recent
admixture---which is also plausible.

\begin{figure}
{\centering\input{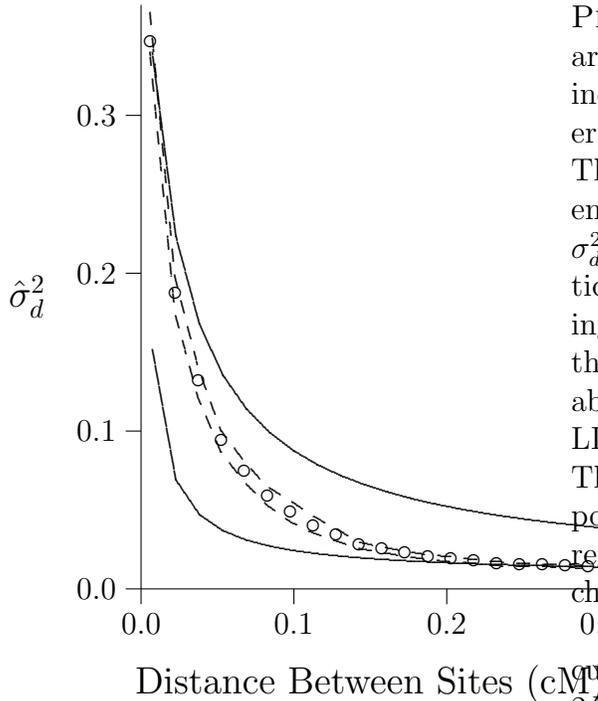}\\}
\caption{\textbf{LD curve for chromosome~1 in Europe.} Open circles
  show the estimated values of $\sigma_d^2$, and dashed lines show a
  95\% bootstrap confidence region, generated by moving-blocks
  bootstrap. Solid lines show the equilibrium curves for $2N=5000$ and
  $2N=30,000$. All analyses are based on a sample of 120
  chromosomes. Source: CEU data set from the 1000-Genomes project
  \citep{Mgenomes:N-491-1}.  Genetic map data, downloaded from the
  1000-Genomes website, were estimated from phased haplotypes in
  HapMap Release 22 (NCBI 36) \citep{HapMap:N-449-851}.}
\label{fig.ceu01}
\end{figure}

In Fig.~\ref{fig.ceu01}, the open circles show $\hat\sigma_d^2$ values
estimated from chromosome~1 in European data
\citep{Mgenomes:N-491-1}. These are surrounded by dashed lines, which
indicate a 95\% confidence region generated by moving blocks
bootstrap. These show that chromosome~1 provides enough data for
accurate estimates of $\sigma_d^2$. In 1981, \citet{Hill:GR-38-209}
expressed skepticism about the possibility of estimating population
size from data on LD. At that time only limited data were available,
and it was not possible to estimate LD statistics with any great
accuracy. This inaccuracy bled into estimates of population size. The
narrow confidence region in Fig.~\ref{fig.ceu01} shows that things
have changed.

The solid lines show equilibrium curves for the cases $2N=5000$ and
$2N=30,000$. Clearly, the observed curve declines more steeply than
either equilibrium curve. As we have seen, this suggests a history of
population expansion in Europe, in agreement with many other analyses
of European data. To go farther than this, we would need to fit
parameters describing population history. Methods for this purpose
will be described in a separate publication.

\section{Discussion}

$\sigma_d^2$ has never been valued for its own sake. It is seen
instead as an approximation to the quantity of real interest---the
expected value of $r^2$. Yet it is often a poor approximation, even in
populations of constant size \citetext{\citealp{Maruyama:MEP-82-151};
  \citealp[pp.~616--617]{Hudson:G-109-611}}. It is even
worse when population size varies.

Following a change in population size, the mean of $r^2$ converges
toward its new equilibrium far faster than does $\sigma_d^2$.
Presumably, this is because $\sigma_d^2$ is sensitive to loci with
high heterozygosity, and the gene trees of such loci are
deep. Whatever the cause, this difference in rates of convergence has
two effects.  First, it makes $\sigma_d^2$ useless as an approximation
to $r^2$ in populations that have varied in size. On the other hand,
it also means that $\sigma_d^2$ itself provides a deep record of
demographic history.

To take advantage of this extended record, one must estimate
$\sigma_d^2$ directly from data. This is easy to do, by using averages
in place of the expectations in Eqn.~\ref{eq.sigdsq}.  With
genome-scale data, such estimates are quite accurate. This provides a
measure of LD that is unique in that one can easily calculate expected
LD from the history of population size. With other measures, such
inferences would require extensive computer simulations.

Following a population expansion, drift is weak and the dynamics of LD
are dominated by recombination. The LD curve begins to decline, and
this decline is fastest in the right-hand portion of the curve, where
recombination rates are highest. Consequently, the curve will be
unusually steep for hundreds of generations following a population
expansion.

Following a population collapse, drift becomes strong and dominates
the dynamics.  It pushes LD upward rapidly and relatively uniformly
throughout the curve. Thus, the LD curve becomes high and flat. This
may explain a phenomenon that puzzled \citet[p.~7]{Pritchard:AJH-69-1}:
\begin{quote}
We have several examples in which large regions exhibit more LD than
would be expected under either a model of constant population size or
a mode with rapid population growth. Yet, at the same time studies of
polymorphism at a small scale reveal less LD than would be
expected. These observations at different scales are hard to
accommodate in a single explanation since factors that increase
long-distance LD will tend to have an even larger effect on closely
linked sites.
\end{quote}
As we have seen, this pattern arises naturally from a recent reduction
in population size.

A bottleneck in population size begins with an episode of population
collapse. At the end of the bottleneck, the LD curve is therefore high
but flat. As population size rises at the end of the bottleneck,
genetic drift becomes weak and recombination grows in importance. The
right portion of the LD curve falls faster than the left, so the curve
becomes steeper. 

In the left portion of the curve, recombination is weak. Genetic drift
is also weak, because of the increase in population size. This allows
mutation to play an important role, and its effect distinguishes the
two forms of expansion: from equilibrium and from a bottleneck. In the
former case, the pre-expansion population had little heterozygosity,
so each mutational increment has a large proportional effect on the
denominator of $\sigma_d^2$. After a bottleneck, the opposite is true,
so the left portion of the LD curve declines more slowly.  The curve
becomes even steeper after a bottleneck than after an expansion from
equilibrium.

In summary, (1)~when recombination is strong relative to drift, LD
declines and the curve becomes steeper; (2)~when drift is strong
relative to recombination, LD rises but the curve stays flat; and
(3)~where drift and recombination are both weak, the rate of decline
in LD decreases with heterogyzosity.

The LD curves of two foraging populations are high and flat,
suggesting a recent collapse of population size.  The European
population, on the other hand, has a curve that is quite steep,
suggesting a history of population expansion. This might reflect the
spread of modern humans into Europe, the spread of farmers during the
Neolithic, or the spread of Indo-European speakers.

\section*{Acknowledgements}

I am grateful for comments from 
R. Bohlender,
E. Cashdan,
R. Gutenkunst,
H. Harpending,
W.G. Hill,
and
J. Seger and for access to computer facilities in the laboratories of
Lynn Jorde and Henry Harpending.

\appendix

\section{Appendix}

\subsection{Hill's model of LD evolution}
\label{sec.hillap}

\citet{Hill:TPB-8-117} incorporates mutation using the model of
infinite alleles. However, most modern work involves either DNA
sequence data, or single nucleotide polymorphisms (SNPs). In these
data, alleles are nucleotide states: A, T, G, or C. Nearly all loci
have just two alleles, so it is not appropriate to assume an infinity
of alleles. Yet as we shall see, Hill's \citeyearpar{Hill:TPB-8-117}
results also apply to a more appropriate mutational model---that of
``infinite sites'' \citep{Kimura:InfiniteSites}, which assumes that
mutation never strikes the same site twice.

Hill begins with a vector of moments, associated with alleles $A_h$,
$A_i$, $B_j$, and $B_k$ at loci $A$ and $B$:
\[
y_{hi,jk} = \left(\begin{array}{c}
E[a_h a_i b_j b_k]\\
E\Bigl[a_h b_jD_{ik} + a_hb_kD_{ij} \hspace{3em}\\
\hspace{3em} \mbox{} + a_ib_jD_{hk} + a_ib_kD_{hj}\Bigr]\\
E[D_{hj}D_{ik} + D_{hk}D_{ij}]
\end{array}\right)
\]
Here $a_i$ and $b_j$ are the frequencies of allele $A_i$ at locus $A$
and allele $B_j$ at locus $B$. The disequilibrium coefficients,
$D_{ij}$, are defined such that $a_i b_j + D_{ij}$ is the frequency of
gamete type $A_iB_j$.  The dynamics of these moments are approximately
linear, after dropping terms in $u^2$, $1/N^2$, and $u/N$, where $u$
is the mutation rate and $N$ the diploid population size. Thus, 
\citet[Eqn.~1]{Hill:TPB-8-117} shows that
\[
y_{hi,jk}(t+1) \approx \mathbf{DRM}y_{hi,jk}(t)
\]
where $\mathbf{D}$, $\mathbf{R}$, and $\mathbf{M}$ are matrices
describing the linear effects of drift, recombination, and mutation.
So far, the model describes only the changes in frequency of existing
alleles. Thus it applies not only to Hill's model of infinite alleles,
but also to the model of infinite sites.

To incorporate mutation, Hill defines a new vector,
\begin{equation}
x = \sum_{h\neq i} \sum_{j\neq k} y_{hi,jk}
\label{eq.x}
\end{equation}
where the sums run across all pairs of alleles at each locus. The
dynamics of this new vector include an additive contribution, $\Delta
x_{\hbox{\scriptsize mut}}$, which represents the effect of mutation to new
alleles: 
\begin{equation}
x(t+1) \approx \mathbf{DRM}x(t) + \Delta x_{\hbox{\scriptsize mut}}(t)
\label{eq.DRMx}
\end{equation}
This equation can be iterated across many generations, and it is easy
to incorporate changes in population size.  At the end of this
process, $\sigma_d^2$ is calculated as $x_3/2x_1$
\citep[p.~124]{Hill:TPB-8-117}.

The first term on the right side of Eqn.~\ref{eq.DRMx} applies equally
to both mutational models. Some reinterpretation is required, however,
to relate the second term to the model of infinite sites. In this new
context, $\Delta x_{\hbox{\scriptsize mut}}$ becomes the contribution
of mutations at other sites in the genome, rather than that from new
alleles at the same pair of sites. The mutational increments are
however identical in the two models, as we shall see.

To see that this is so, let us take a close look at the mutational
contributions to $x$, the vector defined above in
Eqn.~\ref{eq.x}. Under the model of infinite sites, mutation never
strikes the same site twice, so each polymorphic site has exactly two
alleles. In this biallelic context, we can simplify Hill's vector of
moments as
\begin{equation}
h = \left(\begin{array}{c}
E[a(1-a)b(1-b)]\\
E[(1-2a)(1-2b)D]\\
E[D^2]
\end{array}\right)
\end{equation}
Here, $a$ and $b$ are the frequencies of alleles $A_1$ and $B_1$, and
$D$ is the coefficient of linkage disequilibrium. It is defined such
that the $ab + D$ is the frequency of gamete type $A_1B_1$. When each
locus is biallelic, Hill's multiallelic moments reduce to $x_1 =
4h_1$, $x_2=4h_2$, and $x_3=8h_3$ \citep[p.~123]{Hill:TPB-8-117}.

The mutational increments, $\Delta x_{\hbox{\scriptsize mut}}$ and
$\Delta h_{\hbox{\scriptsize mut}}$, both depend on the
heterozygosity, $H = E[2a(1-a)]$.  The dynamics of heterozygosity
under infinite sites are the same as those under infinite alleles
\citep[p.~120]{Hill:TPB-8-117}:
\[
H_{t+1} \approx 2u + (1-1/2N-2u)H_t
\]
The $h_i$ are nonzero only if both loci are polymorphic. On the other
hand, our model assumes that mutation affects only monomorphic
sites. This implies that mutational increments to $h$ occur only at
pairs of sites in which one site is polymorphic and the other
monomorphic. Table~\ref{tab.mut} summarizes the case in which $A$ is
initially polymorphic, and a mutation strikes an initially-monomorphic
locus, $B$.

\begin{table*}
  \caption{Effect of a single mutation at $B$, assuming that $A$ is initially
    polymorphic.} 
\label{tab.mut}
\centering
\begin{tabular}{lccccc}
\hline
 & \multicolumn{4}{c}{Gamete frequencies}\\ \cline{2-5}
                      & $A_1B_1$ & $A_1B_0$ & $A_0B_1$ & $A_0B_0$&$D$\\
\hline
\hline
General               & $w$      & $x$     & $y$      & $z$  & $wz-xy$\\
Before mutation       & 0        & $a$     & 0        & $1-a$& 0\\
$A_1$-linked mutation & $1/2N$   & $a-1/2N$& 0        & $1-a$& $(1-a)/2N$\\
$A_2$-linked mutation & 0        & $a$     & $1/2N$   & $1-a-1/2N$&$-a/2N$\\
\hline
\end{tabular}\\
\end{table*}

Consider first the increment to $h_1=E[a(1-a)b(1-b)]$.  Before
mutation, $h_1 = 0$ because $b=0$. After mutation, $b=1/2N$, so $h_1$
becomes $E[a(1-a)(1/2N-1/4N^2)] \approx H/4N$. This accounts for only
half the effect of mutation, because there are also pairs at which $A$
is monomorphic and $B$ polymorphic. The expected effect of a single
mutation is thus approximately $H/2N$. The expected number of such
mutations per generation is $2Nu$. In aggregate, therefore, the
increment from mutation is $\Delta_{\hbox{\scriptsize mut}} h_1
\approx uH$. 

To derive the mutational increment to $h_2=E[(1-2a)(1-2b)D]$, we
assume as before that $A$ is polymorphic but $B$ is not. Because $D$
is zero when either site is monomorphic, $h_2=0$ before
mutation. After mutation, there are two cases to consider.  With
probability $a$, the mutation falls on an $A_1$-bearing gamete, and
$h_2$ becomes $(1-2a)(1-2/2N)(1-a)/2N$, as shown in
Table~\ref{tab.mut}.  On the other hand, with probability $1-a$ the
mutation falls on an $A_0$-bearing gamete, and $h_2$ becomes
$-(1-2a)(1-2/2N)a/2N$.  In expectation, the new value of $h_2$ is 0.
This result is conditional on a mutation at locus $B$, but an
identical argument applies for mutation at locus $A$. Thus,
$\Delta_{\hbox{\scriptsize mut}} h_2 = 0$.

The third moment is $h_3 = E[D^2]$.  Because $B$ is initially
monomorphic, $D^2=0$ before the mutation.  When a mutation occurs at
$B$, it strikes an $A_1$-bearing gamete with probability $a$ and an
$A_0$-bearing gamete with probability $1-b$. The resulting values of
$D$ are shown in table~\ref{tab.mut}. Squaring these and weighting by
$a$ and $1-a$ gives
\begin{eqnarray*}
\lefteqn{E[a\{ (1-a)^2/4N^2 \} + (1-a)\{ a^2/4N^2 \}]}\hspace{5cm}&&\\
&=& H/8N^2,
\end{eqnarray*}
This is the effect on $D^2$ of a single mutation at $B$. There are
$2Nu$ such mutations per generation, so the expected increment from
mutation at locus $B$ is $2Nu \times H/8N^2 = uH/4N$. Finally,
multiply by 2 to account for cases in which $A$ is initially
monomorphic and $B$ polymorphic. This gives $\Delta_{\hbox{\scriptsize
    mut}} h_3 = uH/2N \approx 0$, ignoring terms of order $u/N$.

In summary, 
\begin{equation}
\Delta_{\hbox{\scriptsize mut}} h \approx [uH, 0, 0],
\label{eq.deltah}
\end{equation}
in agreement with \citet[p.~121]{Hill:TPB-8-117}. This shows that the
mutational increment is the same under the models of infinite sites
and infinite alleles. Because of this equivalence, Hill's results
apply equally to both models of mutation.

\subsection{Estimating Linkage Disequilibrium with 
Partially Phased Diploid  Data}

In 1000-genotypes data, many genotypes are phased but not all.  At
different loci, the unphased genotypes may correspond to different
individuals. Thus, we need a method that can deal unphased genotypes
that are scattered throughout the data matrix.

The symbols $j$, $k$, $l$, and $m$ represent alleles and will always
equal either 0 or 1. I will write phased two-locus genotypes in form
$\gt jklm$, which says that a diploid individual has genotype $jk$ at
locus $A$, and genotype $lm$ at locus $B$. This represents the union
of gametes~$\gam jl$ and~$\gam km$. This genotype is unordered, in
that we cannot distinguish the maternal gamete from the paternal
one. Consequently, there is no distinction between $\gt jklm$ and $\gt
kjml$. When I write genotypes in this form, I imply that linkage phase
is known. In other words, genotypes $\gt jklm$ and $\gt jkml$ are not
equivalent.

Consider a tiny two-locus data set, consisting of three diploid
individuals.
\[
G_1, G_1, G_3 = 
\gt{j_1}{k_1}{l_1}{m_1},
\gt{j_2}{k_2}{l_2}{m_2},
\gt{j_3}{k_3}{l_3}{m_3}
\]
On the right, each row corresponds to a locus, and each column to a
gamete.  To estimate linkage disequilibrium $(D)$, we would need to
calculate $S = \sum_{i=1}^6 x_i y_i$, where $x_i$ is the $i$th value in the
upper row and $y_i$ is the corresponding value in the lower one. This
sum can be written as
\begin{eqnarray*}
S &=& s_1 + s_2 + s_3\\
&=& (j_1l_1 + k_1m_1)
+ (j_2l_2 + k_2m_2)\\
&&\mbox{}
+ (j_3l_3 + k_3m_3)
\end{eqnarray*}
Here, $s_i = j_il_i + k_im_i$ is the contribution of the $i$th diploid
genotype. This calculation requires phased data, and it is the only
step where such data are required in estimating $D$. To cope with
unphased data, we need a method to estimate $s_i$.

Consider the function $s\left(\gt jklm \right) = jl + km$. If at least
one genotype is homozygous then phasing doesn't matter. For example,
$s\left(\gt jkll \right) = s\left(\gt kjll \right)$. We can calculate
$s$ regardless of phasing. The only genotypes that need concern us are
those in which both loci are heterozygous.

As mentioned above, all genic values ($j$, $k$, $l$, and $m$) are
either 0 or 1. Double heterozygotes will look either like $\gt 0101$
or $\gt 0110$. (I ignore the equivalent representations $\gt 1010$ and
$\gt 1001$, because genotypes are unordered.) With unphased data, we
cannot distinguish between these two genotypes. Yet they imply
different values: $s\left(\gt 0101\right) = 1$ but $s\left(\gt
  0110\right) = 0$. For double heterozygotes, I replace $s$ with its
expected value, which equals the probability, $w$, that the genotype
is of form $\gt 0101$ rather than $\gt 0110$.

\begin{table}
  \caption{Gamete types and frequencies. $n_i$ is the number of
    copies of gamate~$i$ in the sample, excluding unphased double
    heterozygotes. $p_i$ is the frequency of that gamete type
    within the population. $a$ is the frequency of allele 1 at locus
    $A$, $b$ the frequency of 1 at $B$, and $D$ the conventional
    coefficient of linkage disequilibrium.}
  \label{tab.gamfrq}
{\centering\begin{tabular}{ccl}
       & Sample& Population\\
Gamete & count & frequency\\
\hline
\rule{0pt}{3ex}$\gam 00$ & $n_0$ & $p_0 = (1-a)(1-b) + D$\\[1ex]
$\gam 01$ & $n_1$ & $p_1 = (1-a)b - D$\\[1ex]
$\gam 10$ & $n_2$ & $p_2 = a(1-b) - D$\\[1ex]
$\gam 11$ & $n_3$ & $p_3 = ab + D$
\end{tabular}\\}
\end{table}

To calculate this probability, I begin with standard results for the
frequencies of the four gamete types, as shown in
Table~\ref{tab.gamfrq}. Following \citet{Rogers:G-182-839}, I ignore
recombination in the most recent generation. Under random mating,
these gametes form at random to produce two-locus genotypes. The
frequency of genotype $\gt 0101$ among double heterozygotes is thus
\begin{eqnarray}
w &=& \frac{2p_0 p_3}{2p_0 p_3 + 2p_1p_2}\nonumber\\
&=& \frac{D + Z}{D + 2Z}
\label{eq.w}
\end{eqnarray}
where
\begin{eqnarray*}
Z &=& \alpha - \beta D + D^2,\\
\alpha &=& a(1-a)b(1-b), \quad\hbox{and}\\
\beta &=& a+b-2ab.
\end{eqnarray*}

These results can be used to estimate $D$ using the EM algorithm
\citep{Dempster:JRS-39-1}. Let $K$ represent the number of unphased
double heterozygotes. Each of these is of type $\gt 0101$ with
probability $w$ and of type $\gt 0110$ with probability $1-w$. The
expected log likelihood is
\begin{eqnarray*}
E\ln L &=& \sum_{i=0}^3 n_i \ln p_i
+ K[ w(\ln p_0 + \ln p_3)\\
&& \mbox{} + (1-w)(\ln p_1 + \ln p_2) ]\\
&=& (n_0 + Kw) \ln p_0\\
&& \mbox{} + (n_1 + K(1-w)) \ln p_1\\
&& \mbox{} + (n_2 + K(1-w)) \ln p_2\\
&& \mbox{} + (n_3 + Kw) \ln p_3
\end{eqnarray*}
Note that $E\ln L$ depends on $p_0$, $p_1$, $p_2$, and $p_3$, which
are themselves functions of $D$. The maximum-likelihood estimate of
$D$ is the value that maximizes $E\ln L$.

This is a unidimensional maximization problem, because $D$ is the only
unknown. $D$ cannot fall outside the range $[\underline{D},
  \overline{D}]$, where $\underline{D}$ is the maximum of $-a(1-b)$
and $-b(1-a)$, and $\overline{D}$ the minimum of $a(1-b)$ and $b(1-a)$
\citep[p.~55]{Lewontin:G-49-49}. The initial value of $D$ is set
assuming that $w=1/2$. In each iteration, if $d^2E\ln L / d D^2 < 0$,
then the algorithm tries a Newton step
\citep[p.~68]{Hamming:NMS-73}. If the result is within
$[\underline{D}, \overline{D}]$, then the Newton step is
accepted. Otherwise, the algorithm uses a modified version of the
bisect algorithm \citep[p.~62]{Hamming:NMS-73} to move in the uphill
direction, as indicated by the sign of $d E\ln L / d D$.

The modification to bisect allows the algorithm to make large steps
when it seems likely that the optimum is at a boundary. For example,
if $d E\ln L / d D > 0$ at the current value of $D$, then motion is to
the right, toward $\overline{D}$. Before deciding how far to move, the
algorithm checks the sign of the derivative at $\overline{D}$. If both
derivatives are positive, then the algorithm takes a big step,
moving 80\% of the distance from $D$ to $\overline{D}$. But if the two
derivatives have opposite sign, then the algorithm takes a small step,
moving only 50\% of the way to $\overline{D}$. When $d E\ln L / d D <
0$, the algorithm is similar, except that motion is to the left,
toward $\underline{D}$.

\bibliographystyle{genetics}
\bibliography{defs,arrpubs,molrec,tree,mcmc,math,arr}

\end{document}